\documentclass[useAMS,usenatbib]{mn2e}
\usepackage{graphicx}
\usepackage[english]{babel}
\usepackage{indentfirst}
\usepackage{afterpage}
\usepackage{subfigure}

\def\LaTeX{L\kern-.36em\raise.3ex\hbox{a}\kern-.15em
    T\kern-.1667em\lower.7ex\hbox{E}\kern-.125emX}

\newcommand{\be}{\begin{equation}}
\newcommand{\ee}{\end{equation}}
\def\ba{\begin{eqnarray}}
\def\ea{\end{eqnarray}}

\def\msun{M_\odot}

\def\ltsima{$\; \buildrel < \over \sim \;$}
\def\simlt{\lower.5ex\hbox{\ltsima}}
\def\gtsima{$\; \buildrel > \over \sim \;$}
\def\simgt{\lower.5ex\hbox{\gtsima}}
\def\etal{{\it et~al.~}}

\def\red{\textcolor{black}}

\usepackage[dvipsnames]{color}

\title{\red {The simulation of molecular clouds formation in the Milky Way} }
\author[Khoperskov et al.]
  {S.A.~Khoperskov,$^1$ E.O.~Vasiliev,$^2$ A.M. Sobolev,$^3$ A.V.~Khoperskov$^4$\\
  $^1$Institute of Astronomy Russian Academy of Sciences, Pyatnitskaya st., 48, 119017, Moscow, Russia \\
$^2$Institute of Physics, Department of Physics, Southern Federal University, Stachki Ave. 194, 344090 Rostov-on-Don, Russia\\
$^3$Ural Federal University, Lenin ave. 51, Ekaterinburg 620000, Russia\\
$^4$Volgograd State University, Universitetskiy pr., 100, Volgograd 400062, Russia}
\begin{document}
\date{Accepted 3004 December 15.
      Received 2004 December 14;
      in original form 2004 December 31}
\pagerange{\pageref{firstpage}--\pageref{lastpage}}
\pubyear{3004}
\maketitle

\label{firstpage}

\begin{abstract}
Using 3D hydrodynamic calculations we simulate formation of molecular clouds in the Galaxy. The simulations take into
account molecular hydrogen chemical kinetics, cooling and heating processes. Comprehensive gravitational potential
accounts for contributions from the stellar bulge, two and four armed spiral structure, stellar disk, dark halo and
takes into account self-gravitation of the gaseous component. Gas clouds in our model form in the spiral arms due to
shear and wiggle instabilities and turn into molecular clouds after $t\simgt 100$~Myr. At the times $t\sim 100 -
300$~Myr the clouds form hierarchical structures and agglomerations with the sizes of $100$~pc and greater. We analyze
physical properties of the simulated clouds and find that synthetic statistical distributions like mass spectrum,
"mass-size" relation and velocity dispersion are close to those observed in the Galaxy. The synthetic $l-v$ (galactic
longitude - radial velocity) diagram of the simulated molecular gas distribution resembles observed one and displays a
structure with appearance similar to Molecular Ring of the Galaxy. Existence of this structure in our modelling can be
explained by superposition of emission from the galactic bar and the spiral arms at $\sim$3-4~kpc.
\end{abstract}


\section{Introduction}

Molecular clouds represent one of the major constituents of our galaxy \citep{Dame2001}. Their masses and sizes vary
in a wide range \citep[e.g.][]{Solomon1987}. Molecular clouds play very important role in determination of the structure
and evolution of the Galaxy. This can be illustrated by the fact that the stars and their clusters originate from
molecular clouds \citep{Shu1987,McKeeOstriker2007,LadaLada2003}. The dynamical time of disintegration for molecular
clouds in the Milky Way is much shorter than the rotation period of the galaxy \citep[e.g.][]{BlitzShu1980}. So, explanation
of the observational fact of existence of the rich population of molecular clouds requires involvement of efficient
mechanisms leading to formation of molecular clouds in the Galaxy. Special interest is paid to formation of the giant
molecular clouds (GMC) which have masses $\simgt 10^6\msun$ \citep{Dame1986}. Two main theories of the GMC formation
were proposed. The first one explains formation of the GMCs by large scale magnetic and/or gravitational instabilities
\citep{Elmegreen1979,Balbus1985,Elmegreen1994,Chou2000,KimOstriker2006}. In the second one the GMCs are built up by
coalescence of the smaller molecular clouds
\citep{FieldSaslaw1965,LevinsonRoberts1981,Tomisaka1984,RobertsStewart1987,KwanValdes1987} or interaction of the
turbulent flows \citep[e.g.][]{BallesterosParedes2007}. In reality both types of processes are expected to play
significant role \citep{ZhangSong1999,Dobbs2008}.

Large scale perturbations in a gas of the Galactic disk can be generated not only by magnetic fields or self-gravity, but also by the galactic spiral shock wave \citep{Roberts1969,Nelson1977}. A simple analysis of the global galactic shock wave shows that such shocks are unstable \citep{Mishurov1975}. Obviously, large scale shear flows in the Galactic shocks lead to the development of both wiggle and Kelvin-Helmholz instabilities \citep{Wada2001,WadaKoda2004}. Such hydrodynamic instabilities are expected to give a birth to spurs, gaseous fragments and turbulent flows \citep{Wada1994,Wada2000,Dobbs2006,Shetty2006}. The gravitational and thermal (in general, thermo-chemical) instabilities are expected to develop further fragmentation, which leads to formation of the population of molecular clouds.

The properties of molecular clouds (mass, size, density, temperature etc.) are intensively studied both observationally
\citep{Solomon1987,Falgarone1992,Lada2008,Heyer2009,Rathborne2009,Kauffmann2010,RomanZuniga2010} and numerically
\citep{DobbsBonnellPringle2006,Dobbs2007a,Dobbs2007b,TaskerTan2009,GloverMacLow2011}. Efforts in order to understand
dependencies between physical parameters of the clouds lead to establishment of several empirical relationships for the
properties of molecular clouds \citep{Larson1981}. In particular, the mass-size relation reflects the hierarchical
spatial structure of clouds \citep{Beaumont2012}. Molecular clouds can agglomerate in small groups and chains, thus
reflecting processes of the cloud formation: disintegration into smaller clumps or combining into bigger clouds. Due
to the origin of the clouds from the large scale perturbations in the galactic disk molecular clouds can trace the
large-scale structures, e.g. spiral arms and the bar \citep[e.g.][]{WadaKoda2004}. Several observational studies display
existence of the galactic molecular ring \citep{Stecker1975,Cohen1977,Roman-Duval2010}, however it is unclear whether
this is an actual ring or just superposition of emission from molecular clouds belonging to different spiral arms
\citep{DobbsBurkert2012}.

In this paper we develop a model aiming to reproduce general characteristics of the formation of molecular clouds in
the Galaxy, their physical properties, their distribution in the Milky Way using the 3D simulations with the comprehensive
gravitational potential, self-gravity of gaseous component, molecular hydrogen chemical kinetics, cooling and heating
processes.

The paper is organized as follows. Section 2 describes basic premises and equations of our model. Section 3 presents numerical results on the disk evolution. Section 4 presents analysis of the physical properties of the clouds in our model. Section 5 considers large-scale structures in the simulated Galaxy. Section 6 summarizes the results.

\section{Model}

\subsection{Gas dynamics and galaxy potential}
The dynamics of the chemically reacting gas mixture in our model of the Galaxy can be written in the single-fluid approximation as follows:
\begin{equation}
 \label{eq::hydro1}
 \displaystyle \frac{\partial \rho }{\partial t} + \nabla \cdot (\rho \textbf{u} ) = 0,
\end{equation}

\begin{equation}
 \label{eq::hydro2}
 \displaystyle \frac{\partial \rho \chi_i}{\partial t} + \nabla \cdot (\rho \textbf{u} \chi_i) = \rho s_i, \;\;
 i=1,...,n_s,
\end{equation}

\begin{equation}
 \label{eq::hydro3}
 \displaystyle \frac{\partial \textbf{u}}{\partial t} + \nabla \cdot (\textbf{u}\otimes\textbf{u})
   = - \frac{\nabla p}{\rho} -\nabla(\Psi_{ext} + \Psi_g),
\end{equation}

\begin{equation}
 \label{eq::hydro4}
 \displaystyle \frac{\partial E}{\partial t} + \nabla \cdot ([E + p]\textbf{u})
   = -\textbf{u}\cdot\nabla(\Psi_{ext}+\Psi_g)- (\Lambda - \Gamma)\,,
\end{equation}
where $\rho$ is the gas density, $p$ is the gas pressure, $\textbf{u} = (u,v,w)$ is the gas velocity vector,
$n_s$ is the number of components in the gas mixture (in our model $n_s=2$),
$\chi_i = \rho_i/\rho$ is the mass fraction of \textit{i}-th component of gas,
$\sum\limits_{i=1}^{n_s} \chi_i = 1$,
$s_i$ is the formation/destruction rate of \textit{i}-th component,
$\displaystyle E = \rho(e + \frac{\textbf{u}^2}{2})$ is the total energy,
$e$ is the specific internal energy,
$\Lambda$ and $\Gamma$ are the cooling and heating rates, correspondingly,
$\Psi_g$ is the gravitational potential of gas,
$\Psi_{ext}$ is the total external gravitational potential produced by the massive dark halo $\Psi_{halo}$,
the stellar bulge $\Psi_{bulge}$ and the stellar disk $\Psi_{disk}$. The latter includes the potentials
of the spiral structure and the bar.
The potential of the gaseous component is determined by the Poisson equation:
\begin{equation}
 \label{eq::poisson}
 \displaystyle \Delta \Psi_g = 4\pi G \rho\,.
\end{equation}
Hereafter we assume $G=1$. The external gravitational potential can be written as follows:
\begin{equation}
 \label{eq::ext-pot}
 \displaystyle \Psi_{ext} =\Psi_{halo} + \Psi_{bulge} + \Psi_{disk}\,.
\end{equation}

\begin{figure}
  \resizebox{\hsize}{!}{\includegraphics{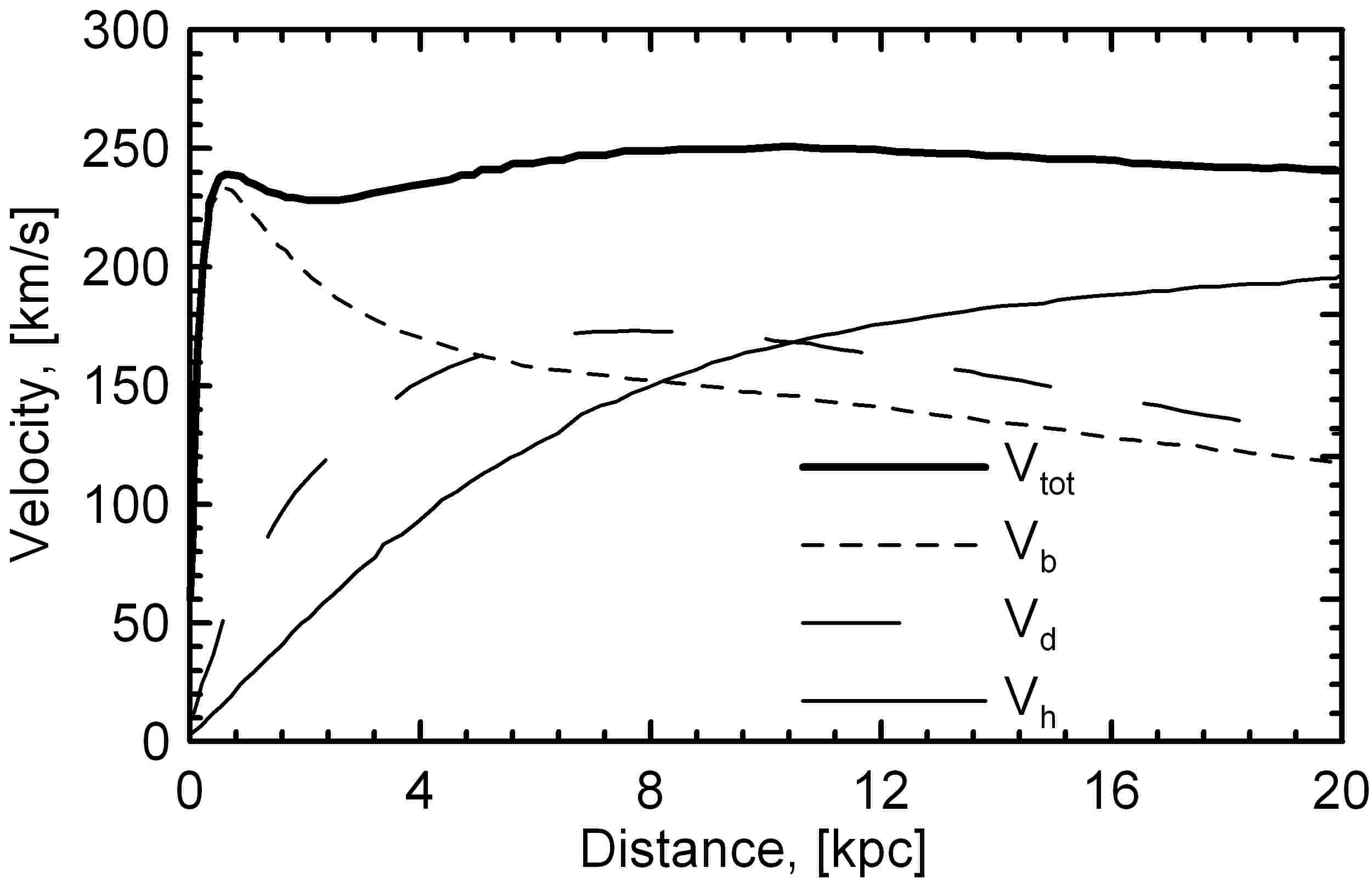}}
\caption{
The rotation curve of the gas components in our model of the Galaxy, $V_{tot}$ (thick solid line).
The contributions from the stellar bulge, $V_b$, the stellar disk, $V_d$, and the dark halo, $V_{h}(r)$,
are shown by dash, long dash and thin solid lines, correspondingly.
}
\label{fig::vrot}
\end{figure}

Figure~\ref{fig::vrot} present the rotation curve of the gas components adopted in our model of the Galaxy.
For the maximum rotation velocity of the Milky Way we adopt a value obtained from the analysis of
the recent observational data on the Galactic masers \citep{Bobylev2010}.
The external gravitational potential of the Galactic subsystems, e.g. stellar bulge, stellar disk and dark halo, $\Phi_{ext}$, produces this rotation curve with the parameters for the Galactic subsystems fitted by \citet{Khoperskov2003}. The detailed description of the gravitational potential components is given in Appendix.

In our calculations we use the value of the angular velocity $\Omega_p = 31$~km~s$^{-1}$~pc$^{-2}$ as a fiducial
value. This value is realized at the radius of corotation $R_c\sim 8$~kpc \citep{Bobylev2010}. Our simulations
have shown that variation of $\Omega_p$ in the range $18-28$~km~s$^{-1}$~pc$^{-2}$
\citep{Amores2009,Sofue2009,Lepine2001} does not produce significant changes in the general properties of evolution
of gas in the simulated Galaxy.

\subsection{Chemical kinetics}

H$_2$ molecules in the interstellar medium are formed on the surface of the dust grains and dissociated
by ultraviolet Lyman-Werner photons and cosmic rays. We suppose that the dust density is proportional
to the gas density because the dust is well mixed with gas
on the scales of our simulations. Thus, evolution of molecular hydrogen number density can be found from \citep{Bergin2004}
\begin{equation}
 \label{eq-h2-formation}
 \displaystyle \frac{dn_2}{dt} = R_{gr}(T)n_1n - [\zeta_{cr}+\zeta_{diss}(N_2, A_V)]n_2,
\end{equation}
where $n_1$, $n_2$ are the number densities of atomic and molecular hydrogen, correspondingly,
$n = n_1 + 2n_2$ is the total number density of hydrogen species, $N_2$ is the H$_2$ column density, $R_{gr}(T)=2.2\times10^{-18}ST^{0.5}$ is the H$_2$ formation rate on dust grains \citep{Tielens1985},
$S = 0.3$ is the efficiency of the H$_2$ formation on dust \citep{Cazaux2004},
$\zeta_{cr} = 6\times 10^{-18}$~s$^{-1}$ is the cosmic ray ionization rate,
$A_V$ is the extinction.
Following \citet{Draine1996} the photodissociation rate can be estimated as:
\begin{equation}
 \label{eq-h2-destr-uv}
 \displaystyle \zeta_{diss}(N({\rm H_2}), A_V) = \zeta_{diss}(0)f_{shield}(N({\rm H_2}))f_{dust},
\end{equation}
where $\zeta_{diss}(0) = 4.17\times 10^{-11}$~sec$^{-1}$ is the unshielded photodissociation
rate, $f_{shield}(N({\rm H_2}))$ is the H$_2$ self-shielding factor, $f_{dust}$ is the dust absorption factor.

To calculate the factors $f_{shield}$ and $f_{dust}$ we need to know both molecular, $N({\rm H_2})$, and total,
$N_{tot} = N({\rm HI}) + 2N({\rm H_2})$, hydrogen column densities. In order to do this accurately we should find a cumulative ultraviolet radiation field produced by the young stars and their clusters.
Obviously, this problem is extremely complex, moreover, star formation processes are not included in our model.
Hence, we follow a simplified approach introduced by \citet{Dobbs2008}. We use a simple estimate that the
column densities of the chemical species are just the local densities of these species multiplied by the typical
distance to a young star, $l_{ph}$: $N({\rm HI,H_2}) = l_{ph} \cdot n({\rm HI,H_2})$. This distance is assumed
to be a constant length scale for the whole disk. In our simulations we take $l_{ph} = 30$~pc, which
is in agreement with the number of massive stars expected from the Salpeter initial mass function.

We assume that the Galactic gas has solar metallicity with the abundances given in
\citet{Asplund2005}: $[{\rm C/H}] = 2.45\times 10^{-4}, [{\rm O/H}] = 4.57\times 10^{-4}, [{\rm Si/H}] =
3.24\times 10^{-5}$. We assume that the dust depletion factors are equal to 0.72, 046 and 0.2 for C, O and Si,
correspondingly. Chemical kinetics of the heavy elements is not solved in our model the. We suppose that the carbon
and silicon are singly ionized and oxygen stays neutral. This assumption is acceptable for the interstellar medium
of the Milky Way due to existence of the strong ultraviolet radiation in the range $\sim$10-13~eV \citep{Habing1968}.

\subsection{Cooling and heating processes}

The cooling rates are computed separately for the temperatures below and above $2\times 10^4$~K. In the low temperature
range the cooling rate $\Lambda$ in the energy equation (\ref{eq::hydro4}) includes the typical processes of radiative
losses for the interstellar medi\-um: cooling due to recombination and collisional excitation and free-free emission
of hydrogen \citep{Cen1992}, molecular hydrogen cooling \citep{Galli1998},
cooling in the fine structure and metastable transitions of carbon, oxygen and silicon
\citep{Hollenbach1989}, energy transfer in collisions with the dust particles \citep{Wolfire2003} and recombination
cooling on the dust \citep{Bakes1994}. In the high temperature range, $T> 2\times 10^4$~K, the cooling rate for
solar metallicity is taken from \citet{Sutherland1993}. We note that in our calculations the gas temperature is
generally below $10^4$~K, higher temperature is reached only in small rarefied regions at the periphery of the disk.

The heating rate $\Gamma$ in the equation (\ref{eq::hydro4}) takes into account photoelectric heating on the dust
particles \citep{Bakes1994,Wolfire2003}, heating due to H$_2$ formation on the dust, and the H$_2$
photodissociation \citep{Hollenbach1979} and the ionization heating by cosmic rays \citep{Goldsmith1978}.

\subsection{Numerical methods}

To solve the system of gas dynamical equations (\ref{eq::hydro1}-\ref{eq::hydro4}) we use nonlinear
finite-volume numerical scheme TVD MUSCL in the Cartesian coordinates \citep{vanLeer1979}. The continuity
equation for chemical species is constructed using the operator splitting scheme. Note that because in our model we have two
chemical species, HI and H$_2$, we can solve the kinetics equation for one of them only, e.g. for H$_2$. At first
we solve the continuity equation without right-hand terms and update the H$_2$ density due to advection. Than
we update this result after solution of chemical reactions. The equation for H$_2$ (\ref{eq-h2-formation}) is integrated using a
fourth-order Runge-Kutta method. To find the gravitational potential of the gas (\ref{eq::poisson}) we use the
TreeCode method \citep{Barnes1986}.

The numerical resolution is $4096\times4096\times20$ cells. The cells are assumed to be cubic with the sides of 7.3~pc.
Therefore, the physical size of the computational grid is $30\times30\times0.15$~kpc. We also made calculations with higher resolution in the vertical direction, but lower in the disk plane direction, e.g. $1024\times1024\times100$, and did not find significant difference in global characteristics of gas evolution. We expect that this is sufficient to model the large scale dynamics of the Galactic disk and to study the formation of the spurs and molecular clouds in the vicinity of the spiral arms.

\section{Disk evolution}

\begin{figure*}
  \resizebox{\hsize}{!}{\includegraphics{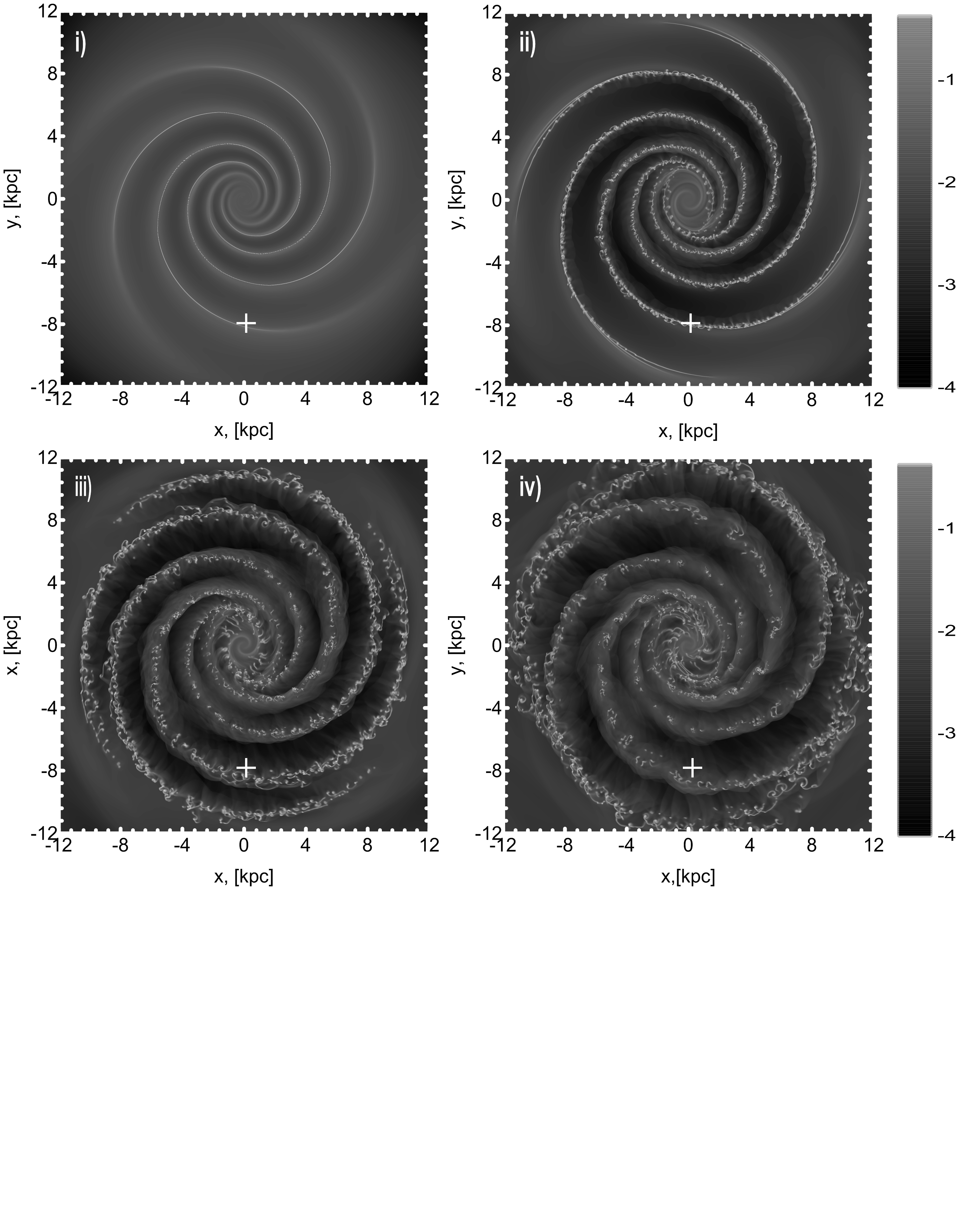}}
\caption{
The evolution of the surface density of gas (g cm$^{-2}$) at $t=50$~Myr (upper left), $t=100$~Myr (upper right),
$t=200$~Myr (lower left) and $t=300$~Myr (lower right). The cross shows position of the Sun (8 kpc).
}
\label{fig::evol}
\end{figure*}

Figure~\ref{fig::evol} shows the surface gas density at 50, 100, 200 and 300~Myr. At the beginning the spiral
shocks are formed due to the supersonic gas flow through the spiral gravitational potential of the stellar disk.
After $t=50$~Myr one finds a well-developed spiral structure in the gas disk (upper left panel). The width of the gas
spirals is significantly smaller than that of the stellar density wave due to higher stellar velocity
dispersion. Owing to shear and wiggle instabilities the shock front becomes perturbed, and as a result the spurs and
fragments are formed. Such spurs and fragments can be found at $t=50$~Myr (upper left panel) and are clearly seen at
$t=100$~Myr (upper right panel). On one hand, higher density in such fragments stimulates molecular hydrogen formation
and, as a consequence, efficient cooling of gas in the fragments. On the other hand, the cooling can lead to further
fragmentation. We find that the efficient molecule formation starts at $t\sim 30-50$~Myr, which corresponds to the
time scale of molecular hydrogen formation on the dust grains. A major part of the H$_2$ molecules is formed in such
fragments, which are the progenitors of molecular clouds. Note that a number of fragments formed may have super-Jeans
masses. So, further fragmentation is expected to be amplified by both thermal instability and self-gravity.

A large number of clouds is formed as a result of further evolution: a flocculent structure in the spiral arms can
be clearly seen at $t=200$~Myr (lower left panel of Figure~\ref{fig::evol}). In the course of collisions the clouds
can merge and form bigger ones or can be disrupted completely. In the latter case they produce a number of smaller
cloudlets. The clouds moving supersonically through the interstellar medium can be stripped due to the Kelvin-Helmholtz
instability. Though H$_2$ molecules are formed and destroyed due to these processes, the total mass of molecular gas
in the Ga\-la\-xy increases until $t\sim 200$~Myr. Then it saturates because the quasi-station\-ary regime is
established between the molecule formation on dust and their destruction by the UV radiation from OB stars (lower
left panel of Figure~\ref{fig::evol}). So, at $t\sim 200$~Myr one finds a population of clouds with sizes varying
in the range $\sim10-100$~pc and typical lifetime about 10~Myr (in the next section we analyze the properties of the
clouds in more detail).

At $t = 200$~Myr groups of clouds are settled along the spiral arms (Figure~\ref{fig::evol}). During further
evolution these groups become larger both in size and mass. Gradually they stand apart from each other in the
spiral arms. Such agglomerations of clouds have sizes more than $100$~pc and the distance between such associations
at $t = 300$~Myr reaches several hundred parsecs. The agglomerations consist of several clouds with the typical
number density $\simgt 1$~cm$^{-3}$. The typical thermal pressure $p_{th}$ inside such clouds is about
2-4~$M_\odot$~pc$^{-2}$~K which corresponds to the temperatures $T\sim 80-200$~K. Actual values of the gas temperature
are lower because considerable portion of the internal energy is spent on the turbulent motions. However numerical
resolution of our simulations is not high enough to resolve the inner structure of the molecular clouds and the
turbulent flows. The turbulence (especially small scale one) represents a common and complex problem. At this stage
we are not able to extract the turbulent part of the internal energy.

\begin{figure}
\resizebox{\hsize}{!}{\includegraphics{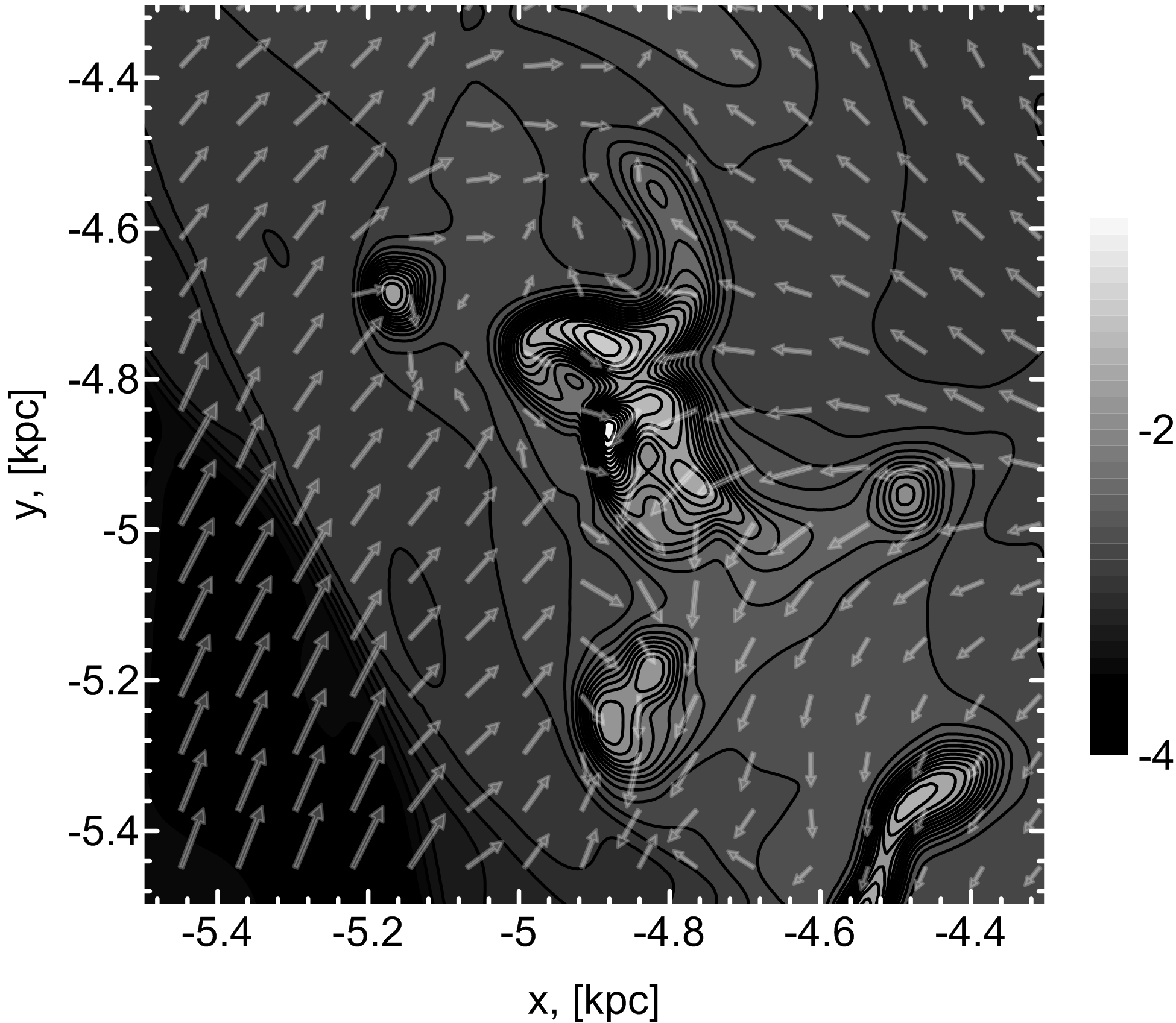}}
\caption{
The surface gas density (g cm$^{-2}$) and the velocity field (in the co-moving coordinate system) of the "cloud"
with coordinates (-4.9~kpc, -4.9~kpc) at $t=300$~Myr.
}
\label{fig::veldisp}
\end{figure}

The velocity dispersion in a gas around molecular clouds is about 3-7~km~s$^{-1}$ and the velocity field around clouds
shows a complex turbulent structure. To consider this we pick out a cloud with coordinates (-4.9~kpc, -4.9~kpc)
at $t=300$~Myr. Figure~\ref{fig::veldisp} shows velocity field around such cloud. The velocities are shown
in the comoving coordinate system connected to the most dense part of the cloud. One can see the irregular
form of the cloud and the colliding gaseous flows in the regions with enhanced density. The structure and physical
parameters of the cloud locations obtained in our numerical simulations are close to those observed in the giant
molecular clouds and their surroundings \citep{Roman-Duval2010}.

\begin{figure}
  \resizebox{\hsize}{!}{\includegraphics{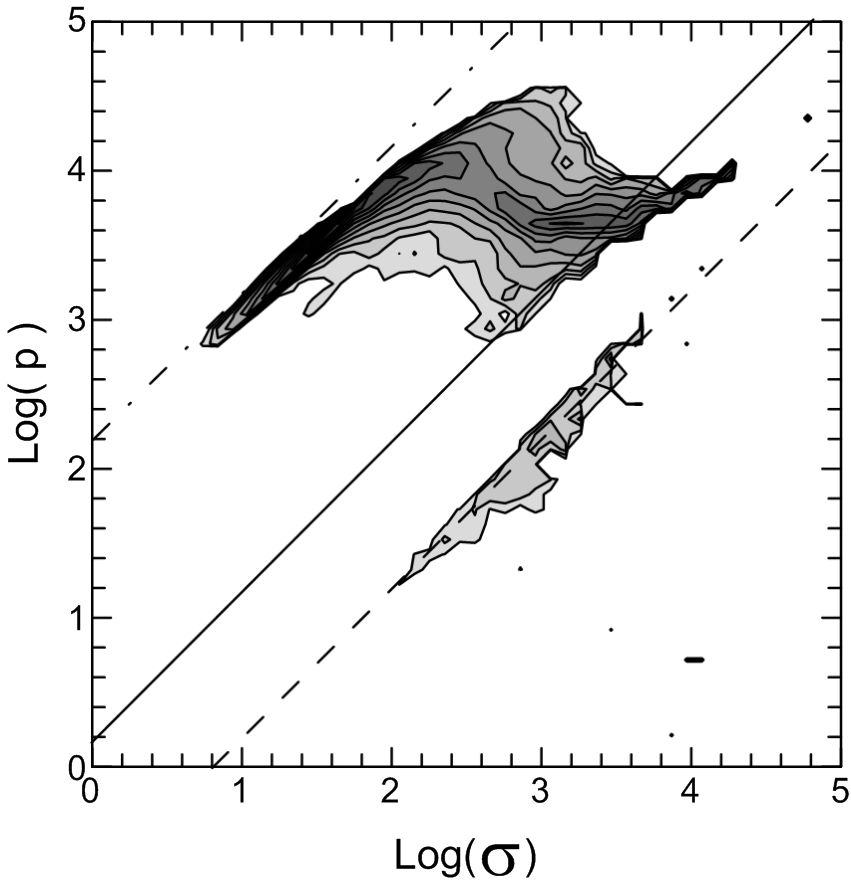}}
\caption{
The phase diagram of thermal pressure $p_{th}$ ($M_\odot$K pc$^{-2}$) versus surface density $\sigma$
($M_\odot$ pc$^{-2}$) at the time $t=300$~Myr. The lines correspond to the constant temperature (see note on the value of the temperature in the text): $10$~K (dashed line),
$10^2$~K (solid line) and $10^4$~K (dash-dotted line).
}
\label{fig::phase}
\end{figure}

Thermo-chemical processes taken into account in our simulations lead to formation of the two phases of the simulated
gas with stable points around $\sim 100$~K and $\sim 10^4$~K (Figure~\ref{fig::phase}, see note on the value of the
temperature above)). A gas with $T\sim 100$~K corresponds to the dense molecular clouds, i.e they have both high number
density ($n\sim 100-300$~cm$^{-3}$) and high H$_2$ molecule abundance ($x({\rm H_2}) \sim 0.3-0.5$). These clouds
concentrate to the spiral arms. A gas belonging to the warmer phase with $T\sim 10^4$~K is a rarefied atomic gas which resides mainly between the arms. A gas of the coldest phase ($T \sim 10$ K) occupies negligibly small volume of space and is not considered here.

\section{The physical properties of \red{molecular} clouds}

In this section we consider physical and statistical properties of the molecular gas in our simulations. This gas
resides in the clumps which are called molecular clouds. A border of the cloud can be defined as location where
H$_2$ abundance or H$_2$ surface density exceeds given limit. The clouds in our analysis are delimited using a
surface density threshold: over the 2D computational grid of the Galactic plane we find isolated groups of cells
with the column density exceeding given threshold. Such clumps normally have irregular form, and their effective
linear size is estimated as a square root of their surface $A$.

For our analysis we have chosen two values of the molecular hydrogen surface density threshold:
\red{ $\Sigma_t^{\rm H_2} =1.5\times 10^{-5}$~g~cm$^{-2}$ and $1.5\times 10^{-4}$~g~cm$^{-2}$. }
 In both cases average density inside a
cloud is greater than several dozens particles per cc, which is close to the typical value of H$_2$ number density
obtained from observations of molecular clouds in the Galaxy \citep[e.g.][]{Shu1987}.

We are using two thresholds for H$_2$ surface density at the border of the cloud in order to show that our basic
conclusions on the statistical properties of clouds are independent of exact value of the threshold. This is important
because there is no single value of H$_2$ surface density which exactly reproduces definition of a molecular cloud
in observations.

\begin{figure}
  \resizebox{\hsize}{!}{\includegraphics{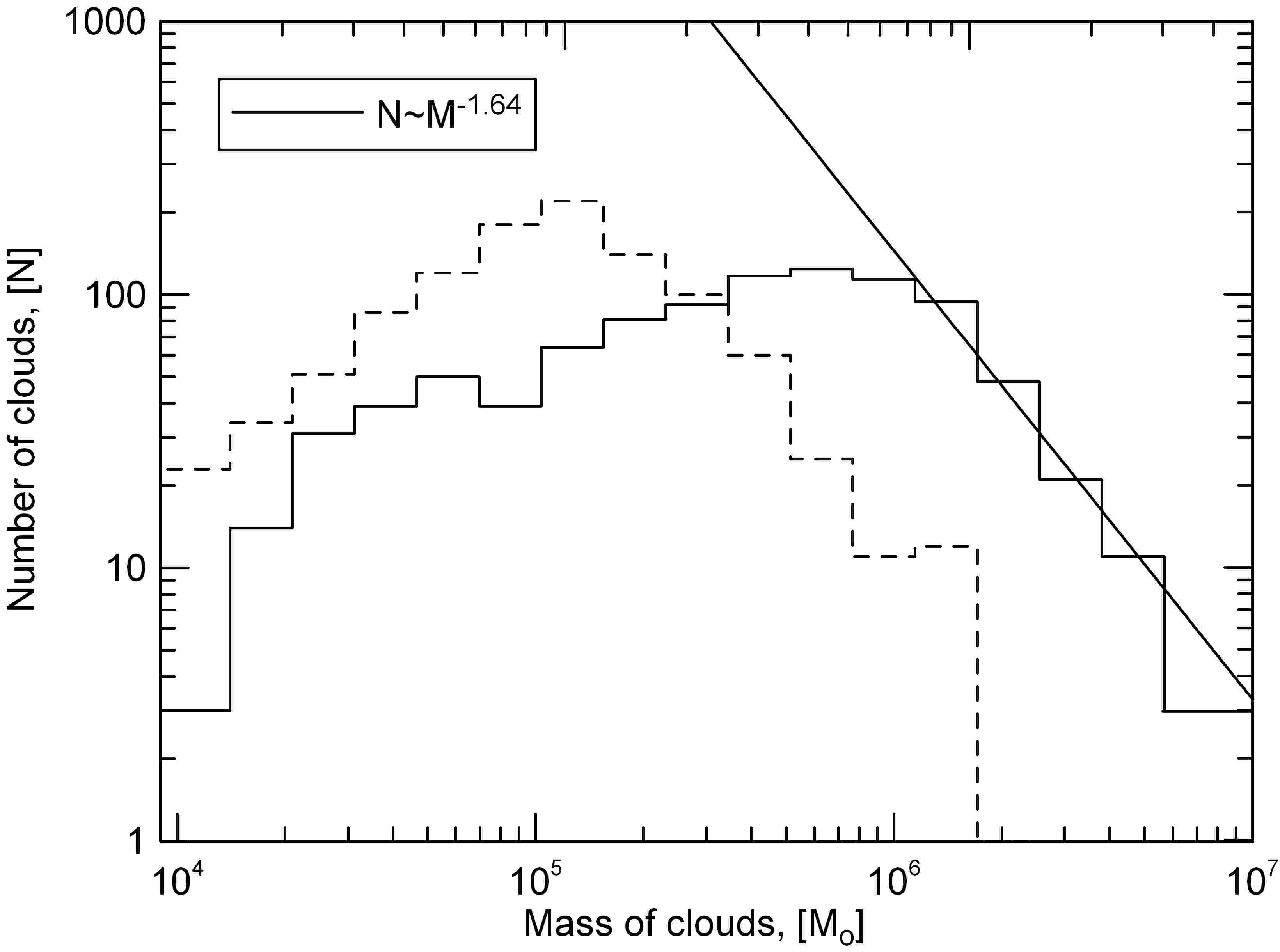}}
\caption{Distribution of masses of the clouds for the H$_2$ surface density thresholds $\Sigma_t^{\rm H_2} =1.5\times
10^{-5}$~g~cm$^{-2}$ (solid line) and $1.5\times 10^{-4}$~g~cm$^{-2}$ (dash line) at $t=300$~Myr. The line corresponds to
the empirical dependence $N\sim M^{-1.64}$ obtained by \citet{Roman-Duval2010}}
\label{fig::massdistrib}
\end{figure}

\begin{figure}
 \resizebox{\hsize}{!}{\includegraphics{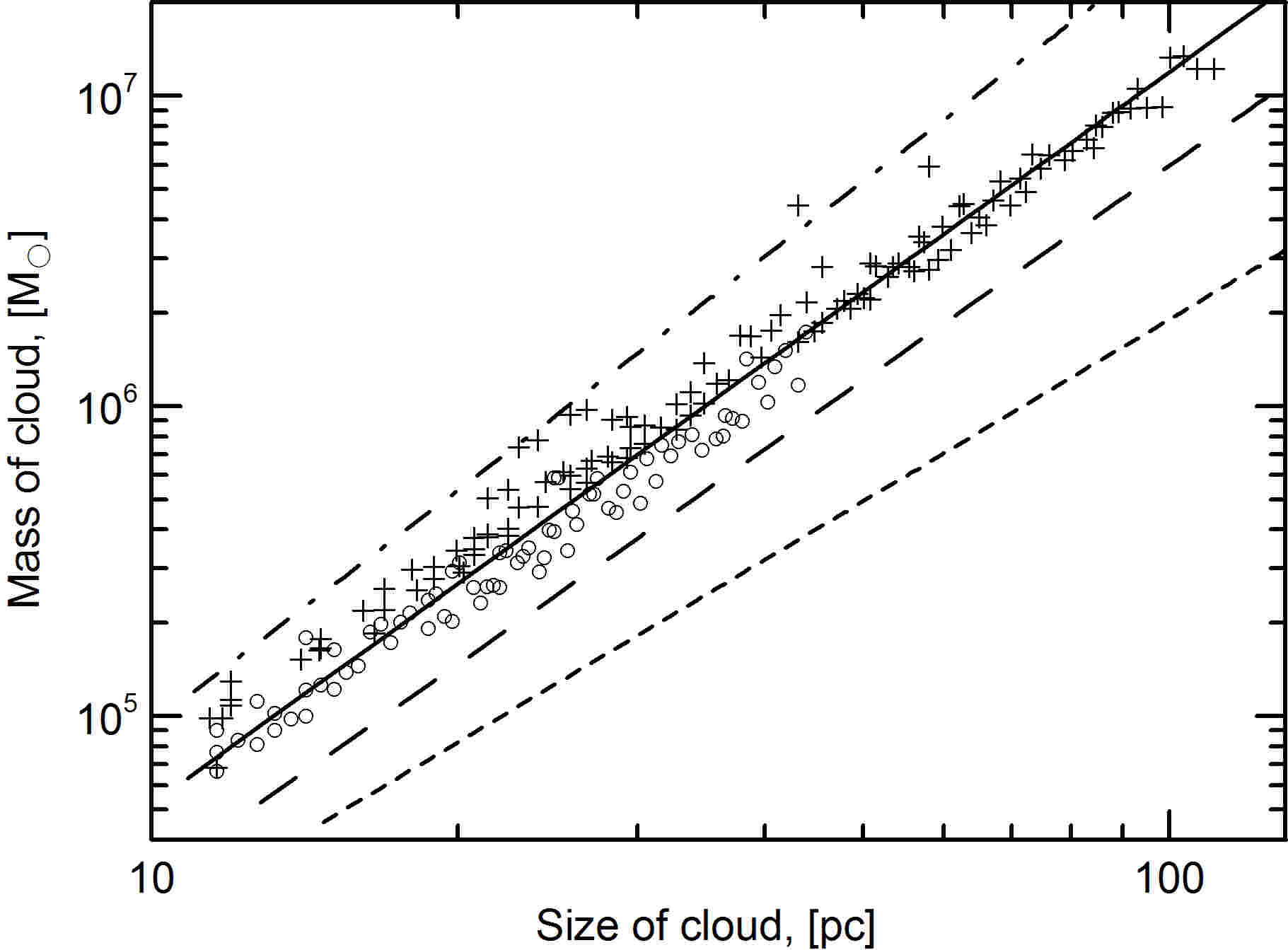}}
\caption{
Mass-size dependence for the H$_2$ surface density thresholds $\Sigma_t^{\rm H_2} =1.5\times 10^{-5}$~g~cm$^{-2}$
(crosses) and $1.5\times 10^{-4}$~g~cm$^{-2}$ (circles) at $t=300$~Myr. The lines correspond to several theoretical and
empirical fits $(M/\msun) = \gamma (A/pc^2)^\beta$ given in \citet{Beaumont2012}:
$(\gamma,\beta) = 240, 0.95$ \citep{Larson1981},
$(\gamma,\beta) = 300, 1.5$ \citep{ElmegreenFalgarone1996},
$(\gamma,\beta) = 150, 1.3$ \citep{RomanZuniga2010} and
$(\gamma,\beta) = 228, 1.36$ \citep{Roman-Duval2010} shown by dotted, dashed, solid and dash-dotted lines,
correspondingly.}
\label{fig::mr}
\end{figure}


Figure~\ref{fig::massdistrib} presents the distribution of masses of the clouds for the H$_2$ surface density thresholds
$\Sigma_t^{\rm H_2} =1.5\times 10^{-5}$~g~cm$^{-2}$ (solid line) and $1.5\times 10^{-4}$~g~cm$^{-2}$ (dash line) at
$t=300$~Myr.

We find that the spectra of cloud masses for both thresholds follow the power law dependence $N\sim M^{-1.64}$:
in the mass range $M\sim 10^5-10^6~M_\odot$ for the higher threshold value and in mass range $M\sim 10^6-10^7~M_\odot$
for the lower density threshold. This invariance of the spectrum on surface density limit indicates to that physical
processes leading to the formation of the structure with density threshold $\Sigma_t^{\rm H_2} \sim 10^{-5} - 10^{-4}
$~g~cm$^{-2}$ have the same nature. Hence, other statistical properties of the structures should be close.
Note that the dependence $N(M)$ for the higher threshold is valid in the same mass range as that obtained in the
observations of the Galactic molecular clouds \citep{Roman-Duval2010}.
For both thresholds one can see a lack of low-mass clouds. In the case of the higher limit the lack of clouds with
$M\simlt 10^5~M_\odot$ can be explained by our numerical resolution, whereas similar deficiency in the observations comes from the incompleteness of the data \citep{Roman-Duval2010}. For the lower threshold the lack of clouds with
$M\simlt 10^6~M_\odot$ can be explained by clustering of smaller and denser clumps. In other words, choosing lower
surface density level we pick out more extended clusters of clumps, which contain both smaller and denser clumps.
In the case of the lower threshold we cover larger regions with lower density, so that the increase of density limit
should lead to decrease of size of the cluster and excludes extended low-density regions from consideration of.

Figure~\ref{fig::mr} presents the dependence of mass of the cloud on its size (mass-size relationship) for the
H$_2$ surface density thresholds $\Sigma_t^{\rm H_2} =1.5\times 10^{-5}$~g~cm$^{-2}$ (crosses) and $1.5\times
10^{-4}$~g~cm$^{-2}$ (circles) at $t=300$~Myr. The sizes of the clouds vary between $\sim 10-60$~pc for the higher
density threshold. The most massive clouds have masses about $2\times 10^6~M_\odot$. These values are in good
agreement with the observational data \citep{Roman-Duval2010}. In some cases the sizes of molecular clouds can be
as large as $\sim 100-200$~pc \citep{Kirsanova2008}, but these regions can be considered as a tight group of smaller
clouds. In our analysis such extended clouds can be found under the lower density threshold (crosses in
Figure~\ref{fig::mr}): there are groups or chains of clouds seen in the periphery of the Galaxy at $300$~Myr
(Figure~\ref{fig::evol}).

The mass-size (or mass-area) relationship among molecular clouds, $M\sim R^2$, which is well-known as
Larson's third law, appears due to cloud evolution and numerous physical processes in the interstellar medium.
Many efforts to obtain or explain this universal relation are given in \citet{Beaumont2012}, who tried to
interpret the relationship in terms of the column density distribution function. However, the scatter of
the empirical fits is large \citep[see lines in Figure~\ref{fig::mr}, note that we pick out only several
fits from the Table~1 presented in][]{Beaumont2012}. One can see that the properties of clouds obtained in the
numerical simulations are close to the power-law relation between the radius and mass of the cloud $M = 228 R^{2.36}$,
which was obtained by applying a $\chi$-square minimization for the observed radii and masses of the Galactic molecular
clouds \citep{Roman-Duval2010}. There is a significant dispersion of our simulated data around this fit.
We find that our best fit of the simulated data $M \sim R^{2.14-2.16}$ (the ranges correspond different density
thresholds) is close to that obtained in observations. Note that our fit has a power-law index close to 2 for
both density thresholds considered here.

\red{Finally, we note that} the velocity dispersion for the majority of clouds varies in the range $v_t \sim
1-2$~km~s$^{-1}$ \red{in our numerical simulations, that is a good agreement with the observational data \citep{Roman-Duval2010}. }


Thus, we have found that the physical properties of clo\-uds obtained from our simulation are close to those observed
in the Galaxy. Our conclusions are coinstrained by the numerical resolution, because we can consider clouds
with sizes larger than 10~pc, whereas even the parsec-size clouds are observed \citep{Roman-Duval2010}. Anyhow, our numerical results provide possibility to consider the global proper\-ties of molecular and atomic gas in the
Galaxy.

\section{The $l-v$ diagram}

\subsection{Molecular hydrogen}

Large-scale distribution of molecular gas in the Galaxy can be represented using the 'longitude-velocity' diagram
(henceforth $l-v$ diagram) constricted from observations of the CO(1-0) line emission \citep{Dame2001}. The CO
molecule is usually used as a tracer of molecular gas \citep{MaloneyBlack1988} because emission of much more
abundant H$_2$ molecule cannot be observed in the cold and warm regions. Current model of molecular kinetics in
our simulations does not provide us with possibility to calculate abundance of CO molecule and the distribution of molecular gas is judged from the distribution of H$_2$. Certainly, the CO-H$_2$ conversion factor depends on many
physical parameters \citep{Shetty2011,Feldmann2012}. However, there are no doubts that the general conclusion on
agreement between synthetic and observational distributions of molecular gas in the Galaxy can be drawn from
comparison of $l-v$ diagram for H$_2$ molecule in the model with CO data of \citet{Dame2001}.

\begin{figure}
  \resizebox{\hsize}{!}{\includegraphics{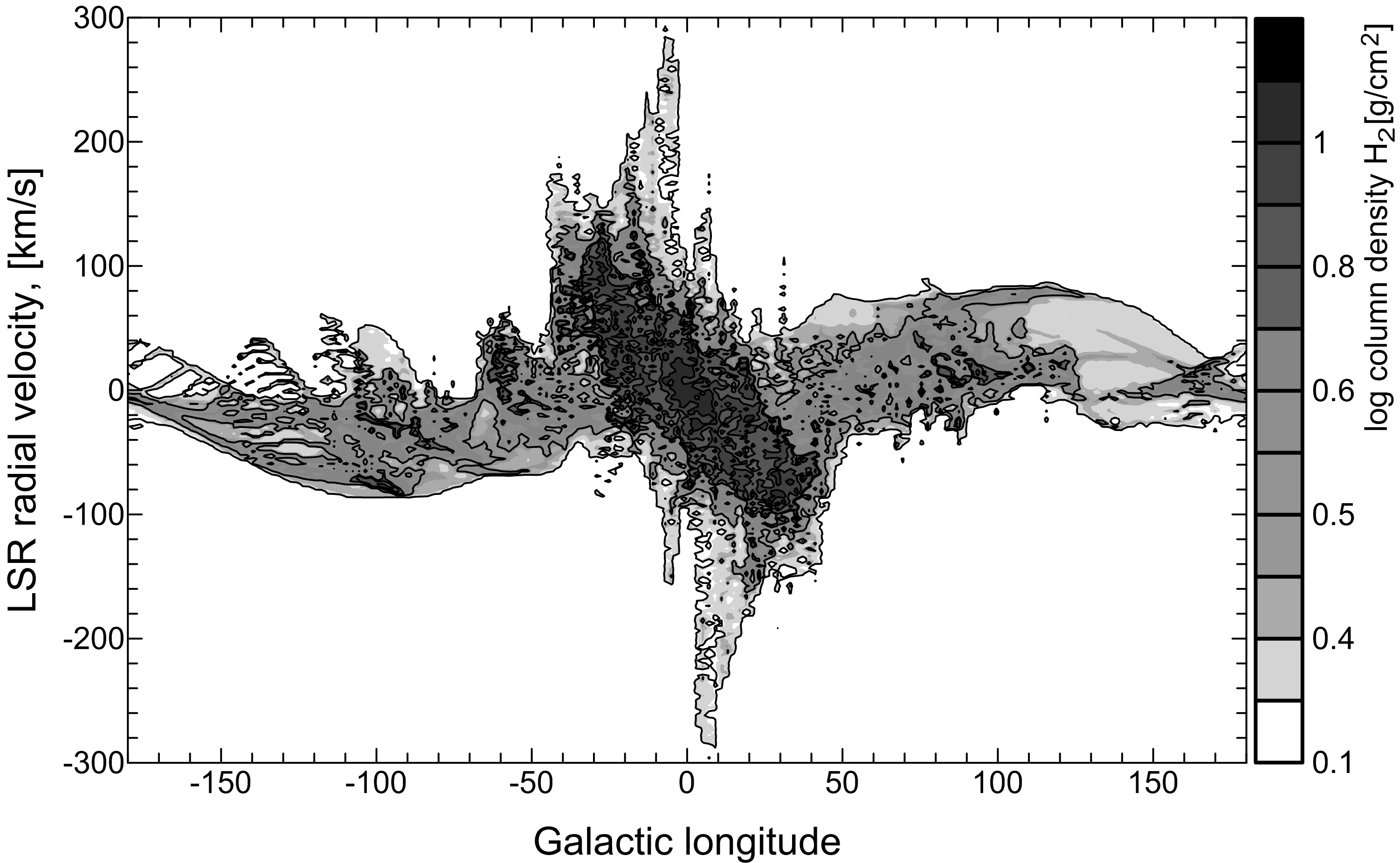}}
\caption{
The \textit{l--v} diagram for the molecular gas at $t=300$~Myr. The mass of gas with a given $(l,v)$ is indicated by
grey scale: from small (light grey) to large (black) mass.
}
\label{fig::diagrH2}
\end{figure}

Figure~\ref{fig::diagrH2} presents the $l-v$ diagram for molecular gas at $t=300$~Myr. The diagram is constructed
for an observer located at the Sun position (see Figure~\ref{fig::evol}). One can find the large-scale structures
corresponding to Perseus, Outer and Carina spiral arms. In the central part of the diagram, $l\pm (30-40)^\circ$,
an intense and extended structure with the strong velocity gradient that passes through the origin is clearly seen.
This structure by its location and general appearance resembles so-called Galactic Molecular Ring
\citep{Stecker1975,Cohen1977,Roman-Duval2010}. However, Figure~\ref{fig::evol} shows that there is no pronounced
ring of molecular material in our simulations. In our model the structure in the Molecular Ring locus of points of
the $l-v$ diagram is associated with the Galactic bar and the spiral arms at $\sim$3-4~kpc. Thus, our results are
in agreement with conclusions on the absence of the Molecular Ring drawn from the simple fitting technique
\citet{DobbsBurkert2012} and previous hydrodynamical simulations \citep{Gerhard1999,Combes2008,Baba2010}. Note that
the previous simulations did not include molecular chemical kinetics.

We have to note on the existence of numerous small-scale structures in the diagram. They are obviously associated
with spurs, agglomerations of clouds and individual clouds. But we make no comment on that here because our
consideration is concentrated on the extended structures scales of which greatly exceed our numerical resolution.


\begin{figure}
  \resizebox{\hsize}{!}{\includegraphics{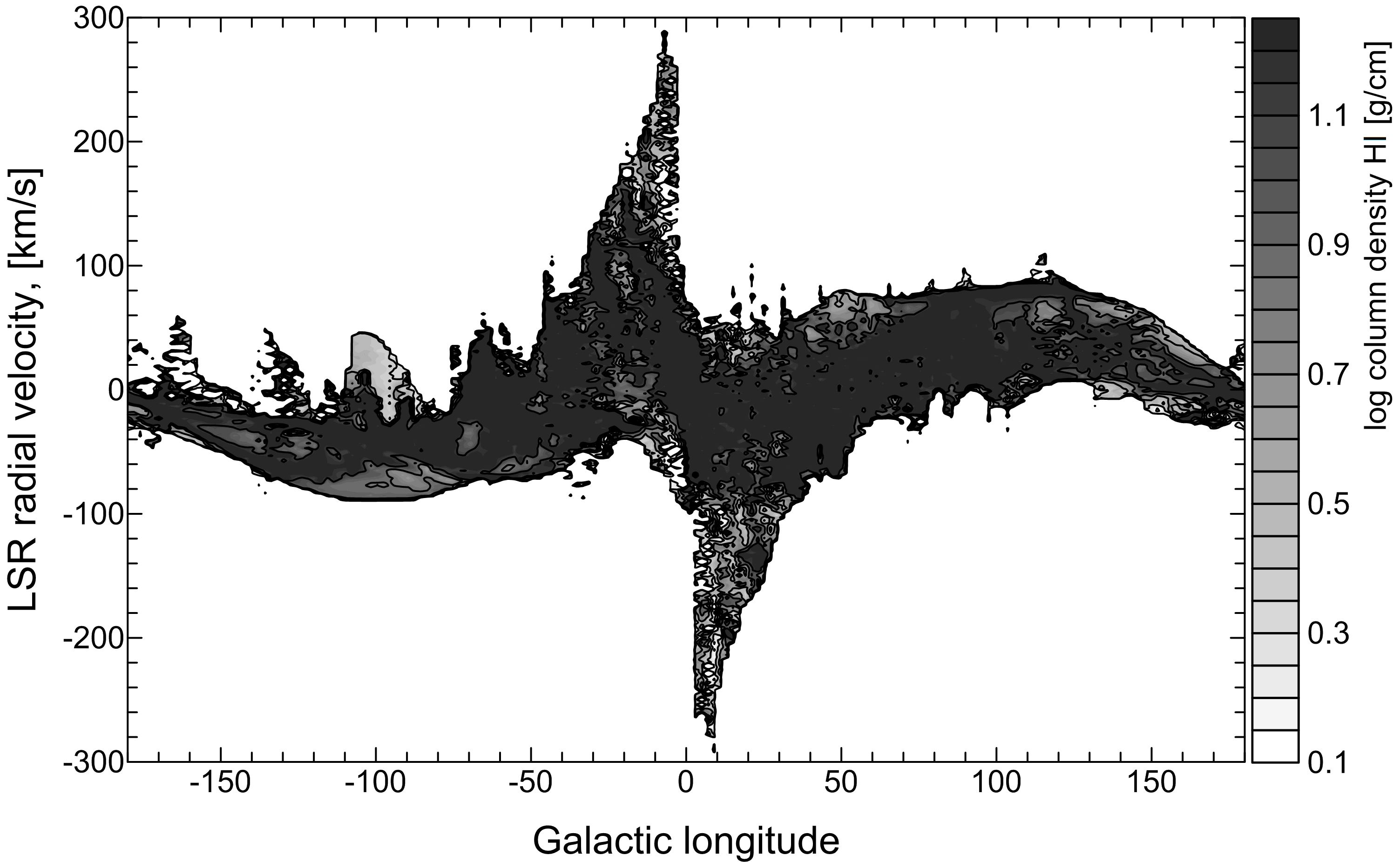}}
\caption{
The \textit{l--v} diagram for the atomic gas at $t=300$~Myr. The mass of gas with a given $(l,v)$ is indicates by grey
scale: from small (light grey) to large (black) mass.
}
\label{fig::diagrHI}
\end{figure}

\subsection{Atomic hydrogen}


The atomic gas in our model is the main part of the gaseous component of the spiral arms and its distribution traces
large-scale structures in the Galaxy. Figure~\ref{fig::diagrHI} presents the synthetic $l-v$ diagram for atomic gas
at $t=300$~Myr. The diagram is constructed for an observer located at the Sun position (see Figure~\ref{fig::evol}).
The structures corresponding to the spiral arms are clearly seen in this Figure. The synthetic diagram does not show
any prominent structure at the locus of points corresponding to the Galactic Molecular Ring Figure~\ref{fig::diagrHI}).
Results of our modelling show that the atomic gas in the simulated Galaxy is distributed much more uniformly than the
molecular gas. The synthetic $l-v$ diagram shows reasonably good agreement with the observational data \citep[see, e.g.
$l-v$ diagrams in][]{kalberla08,mcclure04}. So, our model reproduces main features of the observed distribution of the
neutral gas in the Galaxy.

\section{Conclusion}

In this paper we have studied formation of gas clouds in the model of our Galaxy. The 3D simulations have taken into account molecular hydrogen chemical kinetics, cooling and heating processes. Comprehensive gravitational potential have been used. It included contributions of self-gravitating gaseous component, stellar bulge, two and four armed spiral structure, stellar disk and dark halo. We have analyzed general properties of the simulated clouds and have compared them with statistical distributions taken from the recent surveys of molecular clouds \citep{Roman-Duval2010}.

Our results can be summarized as follows:
\begin{itemize}
 \item the following stages of evolutionary sequence of the molecular cloud system formation were distinguished in
  our modelling: a) spurs and fragments (progenitors of molecular clouds) are formed in the spiral arms due to shear
  and wiggle instabilities at the times $t\simlt 50$~Myr; b) numerous molecular clouds start to form after $t\simgt
  50$~Myr which corresponds to the time scale of molecular hydrogen formation on dust grains; c) a well-developed
  hierarchical structure of the clouds is formed at $t\sim 200$~Myr; d) at the moment of time $t\sim 300$~Myr the
  clouds agglomerate into extended associations with sizes exceeding $100$~pc while the sizes of individual clouds
  vary in the range $\sim10-100$~pc;
 \item the total mass of molecules in the Galaxy increases until $t\sim 200$~Myr, when it saturates; the major part
  of molecule hydrogen is locked in molecular clouds where the H$_2$ molecule abundance reaches $x({\rm H_2}) \sim
  0.3-0.5$;
 \item the statistical dependencies, e.g. the mass spectrum of clouds, the "mass-size" relation and the velocity
  dispersion of clouds, obtained in our simulations are close to those observed in the Galaxy \citep{Roman-Duval2010};
 \item the structure is clearly seen in the Galactic Molecular Ring's locus of points on the simulated $l-v$ diagram
  of molecular gas; this structure doesn't correspond to a real ring of molecular gas and likely arises due to
  superposition of emission from the galactic bar and the spiral arms at $\sim$3-4~kpc, which supports conclusion
  made by \citep{DobbsBurkert2012}; no prominent structure in the Galactic Molecular Ring locus of points is
  distinguished in the simulated $l-v$ diagram of the HI gas.
\end{itemize}

\section{Acknowledgements}

\noindent

We thank A.V. Zasov  and Yu.A. Shchekinov for thoughtful comments
on the manuscript.
This work was supported partially by Russian Foundation for Basic Research (grants 11-02-12247-ofi-m-2011,
10-02-00231, 12-02-00685-a) and the President of the Russian Federation grant (SS-3602.2012.2).
S.A.K. and E.O.V. thank to the foundation "Dynasty" (Dmitry Zimin) for financial support.
E.O.V. thanks to Russian Foundation for Basic Research (grant 11-02-01332).
AMS was partly supported by the Russian federal task program "Research and operations on priority directions of development of the science and technology complex of Russia for 2007-2012" (state contract 16.518.11.7074).
The numerical simulations were made on the supercomputer facilities of NIVC MSU "Lomonosov" and
"Chebyshev".


\appendix

\section[]{External gravitational potential model}

The dark matter halo gravitational potential is taken in the quasi-isothermal form \citep{Begeman-etal-1991!rotation-curves-Dark-haloes}:
\begin{equation}
 \label{eq::halo}
 \displaystyle  \Psi_{halo} = \frac{M_h}{C_h} \cdot \left\{\ln(\xi) + \frac{{\rm arctg(\xi)}}{\xi} + \frac{1}{2}
    \ln\frac{1+\xi^2}{\xi^2}  \right\} \,.
\end{equation}
where $\xi = r/a_h$, $M_h$ is the mass of the dark halo within a radius $r_h$, $a_h$ is the scale of the halo and
$C_h = a_h (r_h/a_h - {\rm arctg}(r_h/a_h))$.
In our model we take $r_h=12$~kpc, $a_h=6$~kpc, $M_h = 0.64 \cdot 10^{11}~M_\odot$.

For the stellar bulge potential we adopt the King's model with the cutoff at radius $r_b^{max}$ \citep{FridmanKh2011}
\begin{equation}
\label{eq::bulge}
 \displaystyle \Psi_{bulge} = - \frac{ M_b }{ r C_b } \ln \left[ \frac{r}{r_b} +
       \left(1+\left(\frac{r}{r_b}\right)^2\right) \right],
\end{equation}
with the following parameters $M_b = 0.092 \cdot 10^{11} M_\odot$, $r_b = 0.2$~kpc, $r_b^{max} = 12$~kpc,
where
\begin{eqnarray}
C_b = \ln(r_b^{max}/r_b+\sqrt{1+(r_b^{max}/r_b)^2}) - \nonumber
\\ (r_b^{max}/r_b)/(\sqrt{1+(r_b^{max}/r_b)^2}). \nonumber
\end{eqnarray}

For the three-dimensional stellar disk we take the exponential distribution of surface density, then the
stellar disk potential can be written as follows \citep{BinneyTremaine}
\begin{eqnarray}
 \label{eq::st-disk}
 \displaystyle  \Psi_{disk} = \pi  \sigma_0 z_d \cdot  \ln\left(\cosh\left(z/z_d\right)\right) - \nonumber
     \\- \pi  \sigma_0  r_d \cdot  y \cdot \left( I_0\left(y\right) K_1\left(y\right) -
     I_1\left(y\right)  K_0\left(y\right) \right)\,,
\end{eqnarray}
where $\displaystyle y = \frac{r}{2\cdot r_d}$, $\displaystyle \sigma_0 = \frac{M_d}{ 2 \pi \cdot r_d^2}$ is the central surface density, $M_d = 0.4 \cdot 10^{11} M_\odot$, the radial scale of disk $r_d = 3$~kpc, the vertical
scale $z_d = 100$~pc, $I_0(y), K_0(y), I_1(y), K_1(y)$ are the cylindrical Bessel functions of the first and second
kind, respectively.

Taking account the gravitational potentials of bar and spiral structure of stellar component we can obtain
more realistic distribution of gas in the Galaxy. Following \citet{Cox2002} the perturbed potential of stellar
disk $\Psi_{disk}^*$ can be written in the form of superposition of potentials generated by the bar and the
two and four-armed logarithmic spiral patterns:
\begin{equation}
\label{eq-pot-spiral-arms}
  \displaystyle \Psi_{disk}^* = \Psi_{disk} \cdot (1+\varepsilon(t) \cdot \Psi_{arms}),
\end{equation}
where $\displaystyle \varepsilon(t) $ describes the evolution of the relative amplitude of the spiral stellar
density wave. We assume a linear increase of the amplitude $ \displaystyle  \varepsilon(t<\tau) = 0.1 \cdot
\frac{t}{\tau}$ during $\tau=220$ Myr, after that time the amplitude is constant:
$\displaystyle  \varepsilon(t>\tau) = 0.1$.

The relative perturbation $\Psi_{arms}$ is adopted in the form \citep{WadaKoda2004}:
\begin{equation}
\label{eq-pot-spiral-arms2}
  \displaystyle \Psi_{arms} = \frac{\kappa_1\cos( \Theta_1 )}{\left( 1 + \kappa_1 \right)^{3/2}} +
      \frac{ \kappa_2 \cos( \Theta_2 ) }{ \left( 1 + \kappa_2 \right)^{3/2}}  +
      \frac{\kappa_4 \cos( \Theta_4 )}{ \left( 1 + \kappa_4 \right)^{3/2}}\,,
\end{equation}
where
$\displaystyle \kappa_1 = \left(r/r_{a1}\right)^2$,
$\displaystyle \kappa_2 = \left(r/r_{a2}\right)^2$,
$\displaystyle \kappa_4 = \left(r/r_{a4}\right)^2$;
$r_{a1} = 2$~kpc, $r_{a2} = 7$~kpc, $r_{a4} = 7$~kpc are the radial scales for bar, two-armed and four-armed spiral
patterns, respectively. The functions $\Theta_i$ are:
$$\Theta_1 = 2 \cdot( \phi - f_1 ),$$
$$\Theta_2 = 2 \cdot\left( \phi - f_2 - \cot(i_2) \cdot \ln\left(r/r_{02} \right) \right),$$
$$\Theta_4 = 4 \cdot\left( \phi - f_4 - \cot(i_4) \cdot \ln\left(r/r_{04} \right) \right),$$
$f_1 = f_2 = f_4 = 50^\circ$ are the phases of bar and spiral patterns,
$r_{02} = r_{04} = 3.5$~kpc, $i_2 = i_4 = 18^\circ$ are the scales and the pitch angles of two-armed and four-armed components.

\end{document}